\documentclass[twocolumn,showpacs,amsmath,amssymb]{revtex4}

\usepackage{graphicx}
\usepackage{dcolumn}
\usepackage{bm}
\usepackage{psfig}  
\usepackage{epsfig}

\begin{document}

\title{Resummed QCD Power Corrections to Nuclear Shadowing}

\author{Jianwei Qiu}%

\author{Ivan Vitev}

\affiliation{Department of Physics and Astronomy, 
Iowa State University, Ames, IA 50011, USA }

\begin{abstract}
We calculate and resum a perturbative expansion of nuclear enhanced
power corrections to the structure functions 
measured in deeply inelastic scattering of leptons
on a nuclear target. Our results for the Bjorken $x$-, $Q^2$- and
$A$-dependence of nuclear shadowing in $F_2^A(x,Q^2)$ and the nuclear
modifications to $F_L^A(x,Q^2)$, obtained  in terms of the QCD
factorization approach, are consistent with the existing data.
We demonstrate that the low-$Q^2$ behavior of these data and the
measured large longitudinal structure function point to a critical
role for the power corrections when compared to other theoretical 
approaches. 
\end{abstract}
                                               

\maketitle

In order to understand the overwhelming data from the Relativistic Heavy 
Ion Collider (RHIC) and make predictions for the future Electron Ion 
Collider (EIC) and Large Hadron Collider (LHC) in terms of the 
successful perturbative QCD factorization approach~\cite{Collins:gx} 
we need precise information of the nuclear parton distribution
functions (nPDFs) $\phi^A_f(x,\mu^2)$ of flavor $f$, atomic weight
$A$, momentum fraction~$x$ and factorization scale~$\mu$. 
Although the $\mu$-dependence of leading twist nPDFs can be calculated 
in terms of pQCD evolution equations, the $x$- and $A$-dependent 
boundary condition at a scale $\mu_0$ must be fixed from  
existing measurements, mostly of the structure function $F_2^A(x,\mu^2)$ in 
lepton-nucleus deeply inelastic scattering (DIS)~\cite{Eskola:1998df}.
It was pointed out recently \cite{Frankfurt:2003zd} that 
the available fixed-target data contain significant higher twist
effects hindering the extraction of the nuclear parton distributions.

On the other hand, 
it has been argued that for physical processes where the effective
$x$ is very small and the typical momentum exchange of the collision
$Q \sim \mu$ is not large the number of soft partons in a
nucleus may saturate \cite{Mueller:1999wm}.  
Qualitatively, the unknown boundary of this novel 
regime in $(x,Q)$ is where the conventional perturbative QCD 
factorization approach should fail~\cite{Qiu:2002mh}. 

In this Letter  we present a pQCD calculation of the resummed power 
corrections to the DIS nuclear structure functions and demonstrate 
their importance in extracting nPDFs.  
From our results we {\em quantitatively} 
identify the characteristic scale of these power corrections
$\xi^2\ll m_N^2$, with nucleon mass $m_N=0.94$~GeV. We 
conclude that for $Q^2 \geq m_N^2$  inclusive  lepton-nucleus 
DIS is above the saturation boundary and can be treated systematically 
in the framework of the pQCD factorization approach with resummed 
high twist contributions.


Under the approximation of one-photon exchange, the lepton-hadron
DIS cross section $d\sigma_{\ell h}/dx\, dQ^2 \propto L_{\mu\nu}\, 
W^{\mu\nu}(x,Q^2)$, with Bjorken variable $x=Q^2/(2p\cdot q)$ and
virtual photon's invariant mass $q^2=-Q^2$. The leptonic tensor
$L_{\mu\nu}$ and hadronic tensor $W^{\mu\nu}$  are defined 
in~\cite{Brock:1993sz}. The hadronic tensor can be expressed in
terms of  structure functions based on the polarization states of
the exchange virtual photon: 
$W^{\mu \nu}(x,Q^2) = \epsilon_T^{\mu \nu}\, F_T(x,Q^2) +  
\epsilon_L^{\mu \nu}\, F_L(x,Q^2)$,  where  $\epsilon_L^{\mu \nu}, \,
\epsilon_T^{\mu \nu}$ are given in~\cite{Guo:2001tz}. The transverse 
and longitudinal structure functions are related to the standard 
DIS structure functions $F_1, \, F_2$ as follows: $F_T(x,Q^2) = F_1(x,Q^2)$, 
$F_L(x,Q^2) = F_2(x,Q^2)/(2x) -  F_1(x,Q^2)$ 
if $4 x^2 m_N^2\ll Q^2$~\cite{Brock:1993sz}. In DIS the exchange 
photon  $\gamma^*$ 
of virtuality $Q^2$ and energy $\nu = Q^2/(2 x m_N)$  probes an
effective volume of transverse area $1/Q^2$ and longitudinal extent 
$\Delta z_N \times x_N/x$, where $\Delta z_N$ is the nucleon size, 
$x_N=1/(2 r_0 m_N) \sim 0.1$ 
and $r_0\sim 1.2$~fm~\cite{Qiu:2002mh}. When Bjorken $x \ll x_N $ 
the lepton-nucleus DIS covers  several nucleons in longitudinal 
direction while it is localized in the transverse plane. 
Although a hard interaction involving more
than one nucleon is suppressed by powers of $1/Q^2$~\cite{Qiu:2002mh},
it is amplified by the nuclear size if $x \ll x_N$. 
In the collinear factorization approach
we evaluate the nuclear enhanced power corrections to the structure
functions to any power in the quantity $(\xi^2/Q^2)(A^{1/3}-1)$ 
and keep 
$\xi^2$ to the leading order in $\alpha_s$. 

In terms of collinear QCD factorization, the structure functions at
the lowest order in $\alpha_s$ are given by~\cite{Brock:1993sz}
\begin{eqnarray}
\label{FTLT}
F_T^{\rm (LT)}(x,Q^2) &=& \frac{1}{2} \sum_f Q_f^2\, 
                        \phi_{f}(x,Q^2) 
                   + {\cal O}(\alpha_s) \; ,
\\
F_L^{\rm (LT)}(x,Q^2) &=& {\cal O}(\alpha_s) \;,  
\end{eqnarray}
where  (LT) indicates the leading twist contribution, 
$\sum_f$ runs over the (anti)quark flavors,  
$Q_f$ is their fractional charge and 
$\phi_{f}$ is the leading twist quark distribution: 
\begin{equation}
\phi_f(x,\mu^2) =
\int \frac{d \lambda_0 }{2\pi}\, e^{i x \lambda_0}
\langle p |\bar{\psi}_f(0)\, \frac{\gamma^+}{2p^+}\,  
 \psi_f(\lambda_0) | p \rangle
\label{qDF}
\end{equation}
in the lightcone $A^+ = n^{\mu} A_{\mu}=0$ gauge for hadron momentum 
$p^\mu = p^+ \bar{n}^\mu$, where $\bar{n}^\mu = [1,0,0_\perp]$ and  
$n^\mu = [0,1,0_\perp]$  specify the ``+'' and ``$-$'' lightcone
directions, respectively.  In Eq.~(\ref{qDF}) the 
parameter $\lambda_0=p^+y_0^-$. 


\begin{figure}[t!]
\begin{center} 
\psfig{file=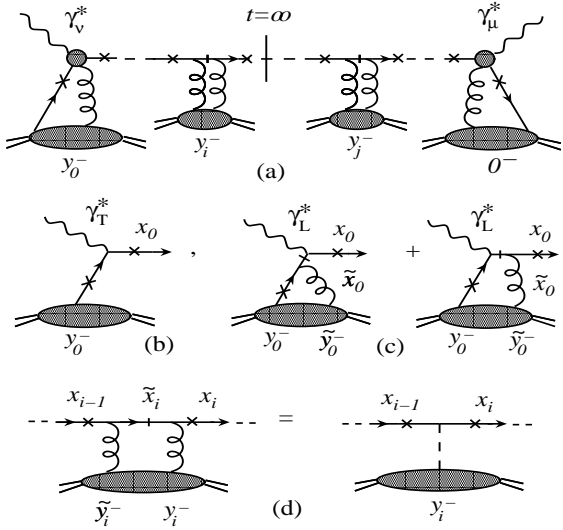,height=2.9in,width=3.in,angle=0}
\vspace*{-.1in}
\caption{(a) Tree level multiple final state scattering of the struck 
quark in DIS on nuclei; 
(b) and (c) leading coupling for transversely and
longitudinally polarized photons to the active partons, respectively; 
(d) effective scalar vertex for the basic two-gluon interactions.}
\label{fig1:F-rule}
\end{center} 
\vspace*{-4mm}
\end{figure}

To compute the nuclear enhanced power corrections
we choose the $A^+=0$ gauge and a frame in which the 
nucleus is moving in the ``+'' direction, $p^\mu \equiv P_A^\mu/A 
= p^+ \bar{n}^\mu$, and the exchange virtual photon momentum is
$q^\mu = -x p^+ \bar{n}^\mu + Q^2/ (2 x p^+) n^\mu$.
In this frame the struck quark propagates along the ``$-$'' direction
and interacts with the ``remnants'' of the nucleus. 
The leading tree level contributions to  $W^{\mu\nu}$ in
lepton-nucleus DIS are given by the Feynman diagrams in 
Fig.~\ref{fig1:F-rule}(a). The cut-line represents the
final state quark  and the blob connected to  $\gamma^*$ 
- the lowest order coupling between the 
photon and the active partons. For transversely polarized photons 
Fig.~\ref{fig1:F-rule}(b)  gives the leading twist 
contribution in Eq.~(\ref{FTLT}).  
For longitudinally polarized $\gamma^*$ the leading tree coupling 
Fig.~\ref{fig1:F-rule}(c) results in the first power correction to $F_L$: 
\begin{eqnarray}
F_L(x,Q^2) &=& F^{\rm (LT)}(x,Q^2) + 
\frac{1}{2}
\sum_f  Q_f^2 \, 4 \left(\frac{4\pi^2 \alpha_s}{3Q^2} \right)  
\nonumber  \\ 
&& \hspace*{-2.cm}    \times \int 
\frac{d \lambda_0 }{2\pi}\, e^{i x \lambda_0} 
\langle p |\bar{\psi}_f(0)\,  \frac{\gamma^+}{2p^+}\,  
 \psi_f(\lambda_0)  \, \hat{F}^2_{\lambda_0} \, | p \rangle  \, ,
\label{FLHT1}
\end{eqnarray}
where the ${\cal O}(\alpha_s)$ $F_L^{\rm (LT)}$ is given in~\cite{FL}  and 
the operator 
\begin{equation}
\hat{F}^2_{\lambda_0} \!   =  \! 
\int \frac{d \tilde{\lambda}_n d \tilde{\lambda}_0 }{2\pi}
\,   \frac{ F^{+ \alpha}(\tilde{\lambda}_n) 
   F_\alpha^{\; +}(\tilde{\lambda}_0) } {(p^+)^2} 
\, \theta(\tilde{\lambda}_{n})
\theta(\tilde{\lambda}_0 - \lambda_0 ) \; .
\label{Fshort}
\end{equation}
Since the effective $\gamma^*_L$ coupling is made of short-distance 
contact terms  of quark propagators~\cite{Qiu:1988dn}, the integrals 
of $\tilde{\lambda}_0$ and $\tilde{\lambda}_{n}$ in Eq.~(\ref{Fshort}) 
should be localized. 


To derive the gauge invariant and nuclear enhanced power corrections at
the tree level, we add gluon interactions to the struck quark and  
convert the gluon field operators in the hadronic matrix element of
$W^{\mu\nu}$ to the corresponding field 
strength~\cite{Qiu:2002mh,Guo:2001tz}. We note that 
there must  be an {\it even} number of interactions 
between the $\gamma^*$  coupling and the final-state cut  because the
quark propagator of momentum $x_ip+q$ has only two types of
terms~\cite{Qiu:1988dn}, $i(\gamma^+/2p^+)/(x_i-x\pm i\epsilon)$ (pole term 
$ \longrightarrow  
\hspace*{-0.6cm} \times \hspace*{0.25cm}$) and
$i(xp^+/Q^2)\gamma^-$ (contact term 
$ \longrightarrow  \hspace*{-0.37cm} \mid \hspace*{0.3cm} $), 
and the gluons are transversely  polarized. 
Consequently, the summation of the coherent multiple scattering 
can be achieved by sequential insertion of 
a basic unit consisting of a pair of gluon 
interactions, left-side of  Fig.~\ref{fig1:F-rule}(d),
connected by a quark contact term, and a quark pole term
to the left (L) or right (R) if the unit
is to the left or right of the final-state cut. 
Integrating over the loop momentum fraction $\tilde{x}_i$, 
we can replace this unit by an effective scalar interaction,  
right-side of Fig.~\ref{fig1:F-rule}(d), with 
a rule,
\begin{equation}
 \left(  x \frac{ 4 \pi^2 \alpha_s}{3Q^2} \right)     
\int \frac{d \lambda_i}{2 \pi} \, \frac{e^{i(x_i-x_{i-1})\lambda_i} }
{x_i - x_{i-1} - i\epsilon} 
\left\{
\begin{array}{ll}
  \frac {\gamma^-  \gamma^+}{2} 
\frac{ -i \hat{F}^2(\lambda_i) }{x_{i-1}-x +i \epsilon }  & {\rm L}   
\\[1ex]
 \frac { \gamma^+  \gamma^-}{2} 
\frac{ -i \hat{F}^2(\lambda_i) }{x_{i}-x -i \epsilon }  & {\rm R}
\end{array}
\right. . 
\label{ScalVert}
\end{equation}
In Eq.~(\ref{ScalVert}) the boost-invariant operator
$\hat{F}^2(\lambda_i)$ is different from 
$\hat{F}^2_{\lambda_0}$ in Eq.~(\ref{Fshort}) and is defined as 
\begin{equation}
\hat{F}^2 (\lambda_i) \equiv  
\int \frac{d  \tilde{\lambda}_i}{2 \pi} \; 
\frac{ F^{+ \alpha}(\lambda_i)  F_{\alpha}^{\; +}(\tilde{\lambda_i}) }
     {(p^+)^2}
\, \theta(\lambda_i  - \tilde{\lambda_i} ) \;.
\label{FFlambda}
\end{equation}
Its expectation value 
can be related to the small-$x$ limit of the gluon distribution,
$\langle p | \hat{F}^2 (\lambda_i) | p \rangle  
\approx \lim_{x \rightarrow 0} \, \frac{1}{2} \, x \, G(x,Q^2)$,
and is independent of $\lambda_i$. It is the 
$\int  d\lambda_i$ in Eq.~(\ref{ScalVert}) that gives the 
$A^{1/3}$-type nuclear enhancement~\cite{Qiu:2002mh}.
Assuming that the interaction leaves the nucleons
in a color singlet state, taking all possible final state 
cuts~\cite{Qiu:2003pm} and performing the integrals over the incoming
partons' momentum fractions we obtain  the   twist $2+2n$ 
contribution to the structure functions:
\vspace*{-.1cm}
\begin{widetext}
\begin{eqnarray}
\delta  F_T^{(n)} \! & \approx & \!  \frac{1}{2} \sum_f Q_f^2 
\left[ x \frac{4\pi^2 \alpha_s}{3 Q^2} \right]^n  
\! \int \frac{d \lambda_0 }{2\pi}\, e^{i x \lambda_0} \, 
\frac{ (i  \lambda_0)^n  }{n!}  \,
\langle P_A |\bar{\psi}_f(0)\,  \frac{\gamma^+}{2p^+}\,  
 \psi_f(\lambda_0) \,   
 \prod\limits_{i=1}^{n} \left[  
\int d {\lambda}_i  \, \theta(\lambda_i) \hat{F}^2(\lambda_i)  \right] 
| P_A \rangle \, ,  
\label{FTmel}   \\
\delta  F_L^{(n)} \! & \approx & \!  \frac{1}{2} \sum_f Q_f^2 
\, \frac{4}{x} \left[ x \frac{4\pi^2 \alpha_s}{3 Q^2} \right]^{n}
\! \int \frac{d \lambda_0 }{2\pi}\, e^{i x \lambda_0} \, 
\frac{(i \lambda_0)^{n-1} }{(n-1)!} \,
\langle P_A |\bar{\psi}_f(0) \,  \frac{\gamma^+}{2p^+} \, 
 \psi_f(\lambda_0) \,  \hat{F}^2_{\lambda_0} \, 
\prod\limits_{i=1}^{n-1} \left[  
\int d {\lambda}_i  \, \theta(\lambda_i) \hat{F}^2(\lambda_i)  \right] 
| P_A \rangle \;, \quad 
\label{FLmel} 
\end{eqnarray}
\end{widetext}
with corrections down by powers of the nuclear size.

We evaluate the multi-field matrix elements in 
Eqs.~(\ref{FTmel}), (\ref{FLmel}) 
in a model of a nucleus of constant lab frame density 
$\rho(r)= 3/(4 \pi r_0^3)$ and approximate the expectation value of 
the product of operators to be a product of expectation
values of the basic operator units in a nucleon state of momentum $p =
P_A/A$:        
\begin{equation}
\langle P_A | \, \hat{O}_0  \, \prod\limits_{i=1}^{n}  
 \hat{O}_i  \,  | P_A \rangle =
A \,\langle p \, | \, \hat{O}_0 \, |  \, p \rangle 
\prod\limits_{i=1}^{n} \left[ N_p \,  
\langle p \, | \, \hat{O}_i \, | \,  p \rangle \right] \;, 
\nonumber 
\end{equation}
with the normalization $N_p = 3/(8 \pi r_0^3 m_N )$. 
The integrals $ \int d\lambda_i \theta(\lambda_i) 
=  (3r_0 m_N/4) (A^{1/3}-1)$
are taken such that the nuclear effect vanishes for $A=1$.  
Resumming the $A^{1/3}$-enhanced power corrections
Eqs.~(\ref{FTmel}), (\ref{FLmel}) we find: 
\begin{eqnarray} 
F_T^A(x,Q^2) \! & \approx & \! \sum_{n=0}^{N}  \frac{A}{n!}
\left[ \frac{\xi^2 (A^{1/3}-1)}{Q^2} \right]^n  x^n \, 
\frac{d^{n}  F_T^{\rm (LT)}(x,Q^2) }{d^n x}   \nonumber  \\
\!&\approx &\! 
A \, F_T^{\rm (LT)}\left( x + \frac{x \xi^2 ( A^{1/3}-1) }{Q^2}, 
                      Q^2 \right) \, ,
\label{FTres}  \\ 
F_L^A(x,Q^2)  \! & \approx & \!  A \, F_L^{\rm (LT)}(x,Q^2)  + 
 \,    \sum_{ n =0}^{N}  \frac{A}{n!} 
\left(\frac{ 4\, \xi^2 }{Q^2} \right)  \nonumber  \\
&& \!\! \times \,   
\left[ \frac{\xi^2 (A^{1/3}-1)}{Q^2} \right]^{n}  x^{n} \, 
\frac{d^{n}  F_T^{\rm (LT)}(x,Q^2) }{d^{n} x}  
 \nonumber \\  
\!  &\approx & \!   
A\,  F_L^{\rm (LT)}(x,Q^2) + 
\frac{ 4\, \xi^2 }{Q^2} \, F_T^A(x,Q^2)  \; ,
\label{FLres}  
\end{eqnarray}
where  $N$ is the upper limit on the number of quark-nucleon interactions  
and $\xi^2$ represents the characteristic scale of  quark-initiated power
corrections to the leading order in $\alpha_s$ 
\begin{eqnarray}
\xi^2  & = &  \frac{3 \pi \alpha_s(Q^2)}{8\, r_0^2} 
\langle p| \, \hat{F}^2 (\lambda_i) \,| p \rangle  \; .
\label{xi2}
\nonumber
\end{eqnarray}
In deriving Eqs.~(\ref{FTres}), (\ref{FLres}) we have taken
$\langle p|\, \hat{F}^2_{\lambda_0}  \,| p \rangle \approx  
(3 r_0 m_N /4) \langle p| \, \hat{F}^2 (\lambda_i) \,| p \rangle$  
and $N\approx \infty$ because the effective value of $\xi^2$ is
relatively small, as shown below.

\begin{figure}[t!]
\begin{center} 
\psfig{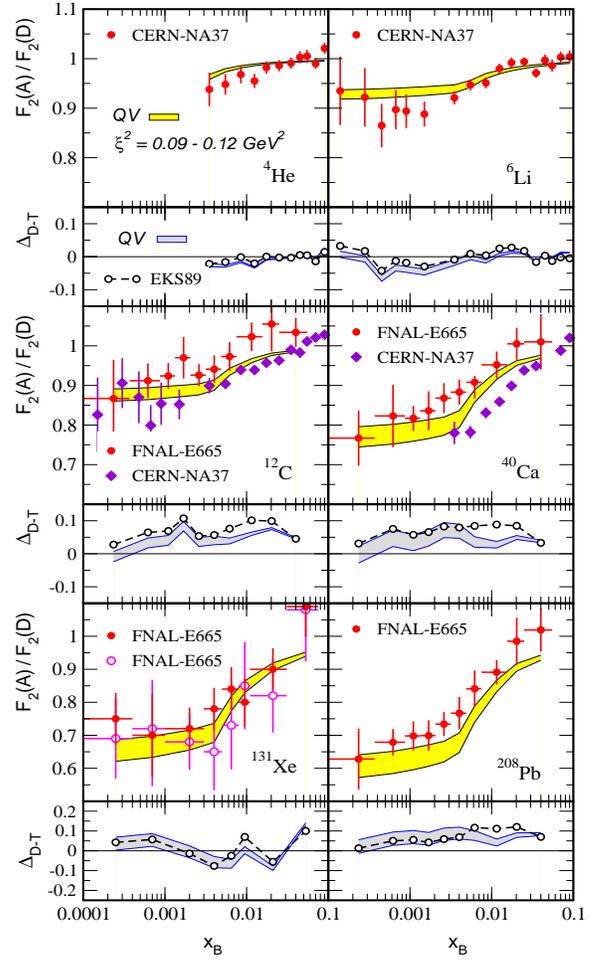}
\vspace*{-0in}
\caption{All-twist resummed $F_2(A)/F_2(D)$ calculation from 
Eqs.~(\ref{FTres}), (\ref{FLres}) versus CERN-NA37~\cite{Arneodo:1995cs} 
and FNAL-E665~\cite{Adams:1995is} 
data on DIS on nuclei. The band corresponds to the choice 
$\xi^2 = 0.09 - 0.12$~GeV$^2$. Data-Theory, where $\Delta_{D-T}$ 
is  computed for the set presented by circles, also shows comparison to the 
EKS98 scale-dependent shadowing parametrization~\cite{Eskola:1998df}.}     
\label{fig3:shad-allT}
\end{center} 
\vspace*{-4mm}
\end{figure}

Eqs.~(\ref{FTres}), (\ref{FLres}) are the main result of this 
Letter. Important applications to other QCD processes and observables 
that naturally follow from this new approach are given 
in~\cite{Qiu:2004qk}.  The overall factor $A$ takes into account 
the leading dependence on the atomic weight and the isospin 
average over the protons and neutrons in the nucleus is implicit.
We emphasize the  simplicity of the end result, which amounts 
to a shift of the Bjorken $x$ by $\Delta x = x \, 
\xi^2(A^{1/3}-1)/Q^2$ with only one parameter 
$\xi^2 \propto \lim_{x\rightarrow0} x G(x,Q^2)$.  
In the following numerical evaluation we use the lowest order 
CTEQ6 PDFs~\cite{Pumplin:2002vw}.


Fig.~\ref{fig3:shad-allT} shows a point by point in ($x,Q^2$)  
calculation of the process dependent modification to $F_2(A)/F_2(D)$ 
(per nucleon) in the shadowing  $x  < 0.1$ region compared 
to NA37 and E665  data~\cite{Arneodo:1995cs,Adams:1995is}.
We find that a value of $\xi^2 = 0.09 - 0.12$~GeV$^2$,
which is compatible with the range from previous
analysis~\cite{Guo:1998rd} of Drell-Yan transverse momentum
broadening  ($\xi^2 \sim 0.04$~GeV$^2$) and momentum imbalance in
dijet photoproduction ($\xi^2 \sim 0.2$~GeV$^2$), makes our
calculations consistent with the both $x$- and $A$-dependence of the
data.  Our calculations might have overestimated the shift in the
region $x$ close to $x_N$ where the $N\approx\infty$ should fail
\cite{Qiu:2004qk}. In Fig.~\ref{fig3:shad-allT}, we impose $Q^2=m_N^2$
for virtualities smaller than the nucleon mass, below which high order 
corrections in $\alpha_s(Q)$ need to be included and the conventional
factorization approach might not be valid. Our result is comparable 
to the EKS98 scale-dependent parametrization~\cite{Eskola:1998df} 
of existing data on the nuclear modification to $F_2^A(x,Q^2)$, 
as seen in the $\Delta_{D-T} = {\rm Data}-{\rm Theory}$ panels of 
Fig.~\ref{fig3:shad-allT}. 
We emphasize, however, that the physical
interpretation is different: in~\cite{Eskola:1998df} the effect is 
attributed to the modification of the input parton distributions at
$\mu_0=1.5$~GeV in a nucleus and its subsequent leading twist scale 
dependence. In contrast, our resummed QCD power corrections 
to the structure functions systematically cover higher twist for
all values of $Q \geq \mu_0$. 
\begin{figure}[t!]  
\begin{center} 
\hspace*{0in} 
\psfig{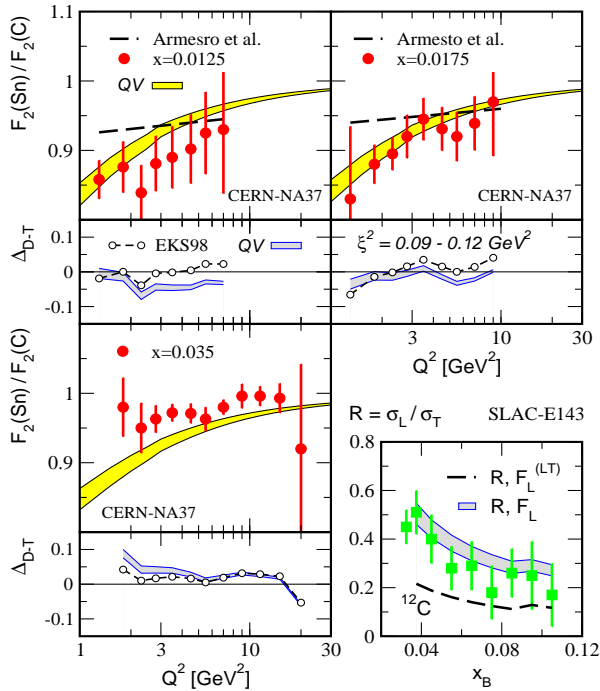}
\caption{CERN-NA37 data~\cite{Arneodo:1996ru} on $F_2(Sn)/F_2(C)$ 
show evidence for a power-law  in $1/Q^2$ behavior consistent with 
the all-twist resummed calculation. Comparison to Gribov-Regge 
model by Armesto {\em et al.}~\cite{Armesto:2003fi} 
and the EKS98 shadowing parametrization~\cite{Eskola:1998df} 
(in $\Delta_{D-T}$) is also 
presented. The bottom right panel illustrates the role of higher twist 
contributions to $F_L$  on the example of $R=\sigma_L/\sigma_T$.  
Data is from SLAC-E143~\cite{Abe:1998ym}.}  
\label{fig4:Q2-allT}
\end{center} 
\vspace*{-4mm}
\end{figure}

With  $\xi^2$ fixed, 
Fig.~\ref{fig4:Q2-allT} shows the  predicted $Q^2$ dependence of 
$F_2(Sn)/F_2(C)$. The $Q^2$ behavior of our result,  distinctly different 
from  a Gribov-Regge  model calculation by Armesto 
{\em et al.}~\cite{Armesto:2003fi}, 
highlights the important role of resummed QCD power corrections,
when presented versus NA37 data~\cite{Arneodo:1996ru}. 
Data-Theory panels again show comparison to the EKS98 
parameterization~\cite{Eskola:1998df}. 
The QCD power corrections at small and moderate $Q^2$ to 
the leading twist structure functions can also be tested via 
the ratio of the cross sections for longitudinally
and transversely polarized virtual photons
\begin{equation}
R(x,Q^2;A)=\frac{\sigma_L}{\sigma_T} = 
\frac{F^A_L(x,Q^2)}{F^A_T(x,Q^2)} \;. 
\end{equation}
Comparison to E143 data for $^{12}$C~\cite{Abe:1998ym} is 
given in the  bottom right panel of Fig.~\ref{fig4:Q2-allT}.  
The dashed curve includes only the bremsstrahlung and 
gluon splitting $F^{\rm (LT)}_L$ from~\cite{FL} 
and is insensitive to modifications of the nuclear parton distributions. 
The gray band represents a calculation of $R$ from 
Eqs.~(\ref{FTres}), (\ref{FLres}).

In conclusion, for $\xi^2=0.09-0.12$~GeV$^2$ our calculated 
and resummed power corrections are consistent with  the 
$x$-,  $Q^2$-  and $A$-dependence of existing 
data  on the small-$x$ nuclear structure functions  without 
any leading twist shadowing.  
Our results, therefore, give  an {\em upper limit} for the 
characteristic scale $\xi^2$ of these power corrections  
in DIS on cold nuclear matter. 
Furthermore, the enhancement of the longitudinal structure function 
$F_L^A(x,Q^2)$ strongly favors a critical role for the higher twist   
in the presently accessible (small $x < 0.1 ,Q^2 \sim\; $few GeV$^2$) 
kinematic regime. 
Less leading twist shadowing and correspondingly less 
antishadowing than currently anticipated will have an important 
impact on the interpretation of the $d+A$ and $Au+Au$ data from 
RHIC~\cite{Vitev:2003xu}.  
The predictions of this systematic approach to the process-dependent 
low $Q^2$ nuclear modification in QCD processes 
will also be soon confronted by copious new data from 
RHIC, EIC and the LHC.

This work is supported in part by the US Department of Energy  
under Grant No. DE-FG02-87ER40371.


\end{document}